\documentclass{article}
\usepackage{amsmath,amssymb}
\usepackage{color}
\begin{document}
\topmargin -1cm

\hfill
\vbox{
  \halign{#\hfil  \cr
	05 April 2007 \cr
}}
\vspace*{4mm} 
\begin{center}
{\large \bf 
CP violation due to multi Froggatt-Nielsen fields
}\\
\vspace*{10mm}
\renewcommand{\thefootnote}{\fnsymbol{footnote}}
{\sc Shinya Kanemura${}^{a,b}$}\footnote{
E-mail: kanemu@sci.u-toyama.ac.jp}, 
{\sc Koichi Matsuda${}^{a,c}$}\footnote{
E-mail: matsuda@mail.tsinghua.edu.cn},  
{\sc Toshihiko Ota${}^{d}$}\footnote{
E-mail: toshi@mpi-hd.mpg.de} \\
{\sc Serguey Petcov${}^{e,f}$}\footnote{
E-mail: petcov@sissa.it}, 
{\sc Tetsuo Shindou${}^{e,f}$}\footnote{
E-mail: shindou@sissa.it}, 
{\sc Eiichi Takasugi${}^{a}$}\footnote{
E-mail: takasugi@het.phys.sci.osaka-u.ac.jp},  
{\sc Koji Tsumura${}^{a}$}\footnote{
E-mail: ko2@het.phys.sci.osaka-u.ac.jp, 
Address after April 2007, Theory Group, KEK} \\ 
\renewcommand{\thefootnote}{\arabic{footnote}}  
\setcounter{footnote}{0}
\vspace*{4mm}
${}^{a)}${\em Department of Physics, Osaka University, 
Toyonaka, Osaka 560-0043, Japan}\\
${}^{b)}${\em Department of Physics, University of Toyama, 
3190 Gofuku, Toyama 930-8555, Japan}\\
${}^{c)}${\em Center for High Energy Physics, Tsinghua 
University, Beijing 100084, China}\\
${}^{d)}${\em Max--Planck--Institut f\"ur Kernphysik, 
Postfach 10 39 80, 69029 Heidelberg, Germany}\\
${}^{e)}${\em Scuola Internazionale Superiore di Studi
Avanzati, I-34014 Trieste, Italy}\\
${}^{f)}${\em Istituto Nazionale di Fisica Nucleare,
Sezione di Trieste, I-34014 Trieste, Italy}\\
\end{center}

\begin{abstract}
We study how to incorporate CP violation in the Froggatt--Nielsen (FN) mechanism. 
To this end, we introduce non-renormalizable interactions with a flavor democratic structure to the fermion mass generation sector. 
It is found that at least two iso-singlet scalar fields with imposed a discrete symmetry are necessary to generate CP violation due to the appearance of the relative phase between their vacuum expectation values. 
In the simplest model, ratios of quark masses and the Cabibbo-Kobayashi-Maskawa (CKM) matrix including the CP violating phase are determined by the CKM element $|V_{us}|$ and the ratio of two vacuum expectation values $R=|R|e^{i\alpha}$ (a magnitude and a phase). 
It is demonstrated how the angles $\phi_i (i=1$-$3)$ of the unitarity triangle and the CKM off-diagonal elements $|V_{ub}|$ and $|V_{cb}|$ are predicted as a function of $|V_{us}|$, $|R|$ and $\alpha$. Although the predicted value of the CP violating phase does not agree with the experimental data within the simplest model, the basic idea of our scenario would be promising to construct a more realistic model of flavor and CP violation. 
\end{abstract}


\section{Introduction}
The Standard Model (SM) of electroweak interactions has been successful. 
It can explain all experimental results except for neutrino oscillation phenomena. Masses of quarks and leptons are generated through the Yukawa interaction after the electroweak symmetry breaking. However, no principle has been established to determine the flavor structure of the Yukawa couplings, and the origin of fermion mass hierarchy remains unknown.

There have been many attempts to explain the flavor structure of Yukawa couplings. A promising approach would be the idea of the flavor symmetry. In models based on the Froggatt--Nielsen (FN) mechanism\cite{FN}, the $U(1)$ global symmetry is imposed as a flavor symmetry, in which the vacuum expectation value (VEV) of an iso-singlet scalar field (FN field) gives a power-like structure of Yukawa couplings due to the $U(1)$ charge assignment for the relevant fields. Extension to more complicated flavor symmetries has also been studied; i.e., non-Abelian global symmetries such as $U(2)$\cite{FS-nonAbelian}, discrete symmetries such as $S_3$\cite{FS-discrete-S3}, $A_4$\cite{FS-discrete-A4}, $D_5$\cite{FS-discrete-D5}, etc. They have several distinct patterns for the symmetry breaking, and the difference in VEVs for each scalar field gives a hierarchical structure of the Yukawa matrix. In most of such models, only the orders of magnitude of the Yukawa matrix elements are estimated, so that $\mathcal{O}(1)$ uncertainties exist in coefficients of the coupling constants between the scalar and matter fields. 
In this framework, CP violation (CPV) comes from complex phases in these coefficients. 

On the other hand, some kinds of the texture such as a democratic structure\cite{Demo} have been investigated for the Yukawa matrix. In the model with the democratic structure, all the elements of the Yukawa matrix are assumed to have the same value up to the leading order, and mass hierarchy and flavor mixing are given by diagonalizing these Yukawa matrices. CPV is supposed to appear as a consequence of complex nature of the terms of tiny breaking of democratic structure. 
In any case the CP violating phase comes from complex phases of free parameters so that it is not predictable.

Since both the flavor mixing and the CP violating phase are determined by the Yukawa matrices, it would be natural to consider that they are given through the same mechanism which is relevant to the Yukawa interaction. In the scenario of spontaneous CPV\cite{SCP}, the phase is deduced from the relative complex phase between VEVs of the scalar fields. Combining the spontaneous CPV scenario with the idea of the flavor symmetry, one can obtain the non-zero complex phase in the Yukawa matrix from the VEVs of scalar fields. This idea has been developed in several flavor models; e.g., a model with three $U(1)$ scalar fields\cite{ABELIAN-HS},  spontaneous CPV in non-Abelian flavor symmetry\cite{SCPV-models}, $SO(10)$ model with the complex VEVs of Higgs field\cite{SO10-with-SCP}, etc. 

In this paper, we introduce a simple model where the FN mechanism works with democratic Yukawa structure between quarks and FN fields, and show how the CPV can be obtained. As mentioned in \cite{ABELIAN-HS}, this type of models requires at least more than two FN fields for a successful prediction of physical CP violating phase. 

This paper is organized as follows. In Sec.II, we study generation of CPV for quarks based on the FN mechanism with the democratic ansatz. In Sec.III, simple models with two FN fields are discussed. We present analytic expressions for the CKM parameters and numerical evaluations are also shown. Conclusions are given in Sec.IV.

\section{CP violation in democratic models}
The democratic ansatz for the flavor structure of the Yukawa matrix has been implemented in Refs.~\cite{Demo,Demo-S3,Demo-geo,Demo-dyn}. In this framework, the Yukawa matrices for the up- and down-type quarks are simply written as
\begin{align}
Y_{u,d}\propto \begin{pmatrix}
1&1&1\\
1&1&1\\
1&1&1
\end{pmatrix}\;.
\label{Demo_Mq}
\end{align}
This flavor structure can be constructed by models with the $S_{3R}\times S_{3L}$ permutation symmetry\cite{Demo-S3}. The symmetry can be realized in the geometrical origin of the brane-world scenario\cite{Demo-geo}, and also in the strong dynamics\cite{Demo-dyn}. Two of the three eigenvalues are zero in these matrices in Eq.~\eqref{Demo_Mq}, and no CP violating phase appears in the $S_{3R}\times S_{3L}$ limit~\footnote{With keeping the $S_{3R}\times S_{3L}$ symmetry, the complex Yukawa matrices $Y_{u,d}$ in Eq.~\eqref{Demo_Mq} can be re-expressed, for example, by an appropriate unitary transformation as 
\begin{align}
Y_{u,d}\propto
\begin{pmatrix}
\omega&\omega^2&1\\
\omega^2&1&\omega\\
1&\omega&\omega^2
\end{pmatrix}\;,
\end{align}
where $\omega=e^{i 2\pi/3}$ is the cube root of one. However the complex phase in this matrix is unphysical, because it is rephased out by the redefinition of quark fields.}. It is clear that in order to explain the experimental data this permutation symmetry must be broken by some small effects. When the small breaking terms for the permutation and CP symmetries are introduced by hand, the mass splitting between the 1st- and 2nd-generation quarks, the mixing angles and the CP violating phase are explained.

The FN mechanism is a simple idea to generate the mass hierarchy of quarks and leptons. In the simplest FN model\cite{FN}, an iso-singlet scalar field, $\Theta$, is introduced with the $U(1)_{\text{FN}}$ flavor symmetry in order to discriminate the fermion flavor by the $U(1)_{\text{FN}}$ charge.

The $U(1)_{\text{FN}}$ charge assigned for $\Theta$ is taken to be $f_\Theta^{}=-1$ without loss of generality. Under the $U(1)_{\text{FN}}$ symmetry, non-renormalizable interactions relevant to the quark mass matrix can be written as
\begin{equation}
\mathcal{L}_{\text{FN}} =
  -(C_U)_{ij} \bar{U}_i Q_j \cdot H_u \left( \frac{\Theta}{\Lambda} 
\right)^{f_{U^c_i} + f_{Q_j} + f_{H_u}}
 -(C_D)_{ij}\bar{D}_i  Q_j \cdot H_d 
\left( \frac{\Theta}{\Lambda} \right)^{f_{D^c_i} + f_{Q_j} + f_{H_d}},
\label{Lagrangian_FN}
\end{equation}
where $H_u$ and $H_d$ are iso-doublet fields (the Higgs fields) with their hypercharge to be $-1/2$ and $+1/2$, respectively\footnote{In the SM, $H_d=i\sigma_2H_u^*$ is satisfied.}, $Q_i$ is the left-handed quark doublet, 
$U_i$ and $D_i$ are right-handed up- and down-type quarks in the $i$-th generation, and $C_U$ and $C_D$ are coupling constants of order one. 
The $U(1)_{\text{FN}}$ charge of the field $X$ is expressed by $f_X^{}$.The cut-off scale is given by $\Lambda$, which describes the mass scale of new physics dynamics. The coefficients $(C_U)_{ij}$ and $(C_D)_{ij}$ are generally complex numbers. 

After $U(1)_{\text{FN}}$ is broken by the VEV of $\Theta$, 
\begin{eqnarray}
\langle\Theta\rangle=\lambda \Lambda \;, 
\end{eqnarray}
where $\lambda$ is a small dimensionless parameter, the quark Yukawa matrices are obtained as
\begin{eqnarray}
(Y_{U,D})_{ij} = 
(C_{U,D})_{ij} \lambda^{f_{U^c_i,D^c_i} + f_{Q_j} + f_{H_{u,d}}}.
\end{eqnarray}
With the assignment of $U(1)_{\text{FN}}$ charges\cite{U1charge} as 
\begin{align}
\left((f_{Q_1},f_{U^c_1}), (f_{Q_2},f_{U^c_2}), (f_{Q_3}, f_{U^c_3})\right)&=(3,2,0)\;,\quad(f_{D^c_1},f_{D^c_2},f_{D^c_3})=(2,1,1),\nonumber\\
(f_{H_u},f_{H_d})&=(0,0),
\label{U1_charge}
\end{align}
observed quark mass hierarchy and the CKM mixings can be derived by assuming $\lambda$ to be close to the Cabibbo angle $\sin\theta_c=0.22$. At the leading order, induced masses for quarks are 
$m_u \sim \lambda^6 \langle H_u\rangle$, 
$m_c \sim \lambda^4 \langle H_u\rangle$, 
$m_t \sim \langle H_u\rangle$, 
$m_d \sim \lambda^5 \langle H_d\rangle$, 
$m_s \sim \lambda^3 \langle H_d\rangle$ 
and $m_b \sim \lambda \langle H_d\rangle $, 
and the CKM matrix is given by 
\begin{equation}
U_{CKM} \sim 
\begin{pmatrix}
1 & \lambda & \lambda^3 \cr
\lambda & 1 & \lambda^2 \cr
\lambda^3 & \lambda^2 & 1\cr
\end{pmatrix}\;.
\end{equation}

Now we consider the possibility of the spontaneous CPV due to the complex phase of $\langle \Theta\rangle$. We assume that $C_U$ and $C_D$ in Eq.~\eqref{Lagrangian_FN} have the democratic structure, {\em i.e.}, 
\begin{align}
C_{U,D}=\alpha_{U,D}\begin{pmatrix}
1&1&1\\
1&1&1\\
1&1&1
\end{pmatrix}\;,
\label{democratic_Y}
\end{align}
There is no CPV in the model with only one FN field. Although complex phases may be obtained in the mass matrices by introducing the complex VEV of $\Theta$, such phases are rotated away by the phase redefinition of quark fields. Hence the model should have at least two FN fields $\Theta_{1,2}$ in order to obtain non-vanishing CP violating phase. We start from the following Lagrangian with two FN fields,
\begin{align}
\mathcal{L}_{\text{FN}_2} =& - \sum_{n_1^u,n_2^u} \bar{U}_i (C_U)_{ij} Q_j \cdot H_u 
\left( \frac{\Theta_1}{\Lambda} \right)^{n_1^u} \left( \frac{\Theta_2}{\Lambda} 
\right)^{n_2^u}
- \sum_{n_1^d,n_2^d} \bar{D}_i (C_D)_{ij} Q_j \cdot H_d 
\left( \frac{\Theta_1}{\Lambda} \right)^{n_1^d} \left( \frac{\Theta_2}{\Lambda} 
\right)^{n_2^d} \;,
\label{FN2_Lag}
\end{align}
where $n_1^{u,d}$ and $n_2^{u,d}$ run from zero to $n_{ij}^{U,D}\equiv f_{U^c_i,D^c_i} + f_{Q_j} + f_{H_{u,d}}$ with satisfying the constraint $n_1^{u,d}+n_2^{u,d}=n_{ij}^{U,D}$. After the $U(1)_{\text{FN}}$ symmetry is broken, the Yukawa couplings in the SM are given by 
\begin{equation}
(Y_{U,D})_{ij} = (C_{U,D})_{ij} \lambda^{n^{U,D}_{ij}}
\sum_{k=0}^{n^{U,D}_{ij}} R^k\;,
\end{equation}
where 
\begin{equation}
\lambda=\frac{\langle\Theta_1\rangle}{\Lambda}\;,\;
R=\frac{\langle\Theta_2\rangle}{\langle\Theta_1\rangle}\equiv |R| e^{i \alpha}\;.
\end{equation}
Therefore, physical CP violating phase can be obtained from the relative phase $\alpha$ between $\langle \Theta_1 \rangle$ and $\langle \Theta_2 \rangle$. 

\section{Examples for the model with two Froggatt-Nielsen fields}

In this section, we show how to generate CPV from two FN fields by considering simple models. In order to generate the quark mass hierarchy, we employ the $U(1)_{\text{FN}}$ charge assignment for matter fields given in Eq.~\eqref{U1_charge}. In general, $U(1)_{\text{FN}}$ charges for the FN fields $\Theta_1$ and $\Theta_2$ can be different with each other. We here assume that the both have the same $U(1)_{\text{FN}}$ charge for simplicity; $(f_{\Theta_1},f_{\Theta_2})=(-1,-1)$.

\vskip 2mm
\noindent
{\em (a) The simplest toy model}

First of all, we discuss the naive model defined in Eq.~\eqref{FN2_Lag}. The mass matrices~$M_{u,d}$ for up- and down-type quarks are given by
\begin{eqnarray}
(M_u)_{ij} = \alpha_U\langle H_u \rangle A_{n_{ij}^U}\lambda^{n_{ij}^U}\;,\quad
(M_d)_{ij} = \alpha_D\langle H_d \rangle A_{n_{ij}^D}\lambda^{n_{ij}^D}\;,
\end{eqnarray}
where $A_n=\sum_{k=0}^{n}R^k$.
This model predicts $m_s^2/m_b^2=\mathcal{O}(\lambda^{6})$ and $|V_{us}|=\mathcal{O}(\lambda)$. Only one of the two experimental values can be adjusted. Furthermore, each mass matrix gives one massless eigenstate because of  the conditions $\det M_u=\det M_d=0$\footnote{Even when the $u$- and $d$-quarks are massless at the electroweak scale, their finite masses would be generated at lower energy scales due to the strong dynamics\cite{massless-up}.}.

In order to avoid these difficulties, the following possibilities can be considered: (i) introducing an additional symmetry, (ii) throwing away the democratic ansatz given in Eq.~\eqref{democratic_Y}, etc. In the next model, we explore the possibility of keeping the democratic structure for $C_{U,D}$. 

\vskip 2mm
\noindent
{\em (b) The extended models with the $Z_2$ symmetry}

We try to construct more realistic model. The $U(1)_{\text{FN}}$ charges are assigned again as in Eq.~\eqref{U1_charge}. In order to obtain the observed value of $m_s^2/m_b^2$ by setting $\lambda\sim \sin\theta_c$, we impose the $Z_2$ symmetry under the transformation of $\Theta_1 \to \Theta_1$ and $\Theta_2 \to -\Theta_2$. For $Z_2$ parity assignment for quark fields, there are a lot of choices. If we consider the scenario associated with the grand unified theory (GUT), it would be natural that the $Z_2$ parity for $Q_i$ and $U^c_i$ is common and that for $H_u$ is set to be $+$ for the prediction of a large top-quark mass. Because $H_d$ always couples to $D^c_i$, we can set the $Z_2$ parity for $H_d$ to be $+$ without loss of generality. Then, there are $64$ possibilities on parity assignment for quarks. However, it turns out that most of them cannot give correct numbers of $m_c^2/m_t^2$ and $m_s^2/m_b^2$. Consequently, only the following sets of $Z_2$ parity assignment are enough to be discussed; 
\begin{itemize}
\item Type I-a,
\begin{align}
\left((Q_1,U^c_1),(Q_2,U^c_2),(Q_3,U^c_3)\right)&=
(+,+,-)\;,\quad
(D^c_1,D^c_2,D^c_3)=
(+,+,-)\;.
\end{align}
\item Type I-b,
\begin{align}
\left((Q_1,U^c_1),(Q_2,U^c_2),(Q_3,U^c_3)\right)&=
(+,+,-)\;,\quad
(D^c_1,D^c_2,D^c_3)=
(-,+,-)\;.
\end{align}
\item Type II-a,
\begin{align}
\left((Q_1,U^c_1),(Q_2,U^c_2),(Q_3,U^c_3)\right)&=
(+,-,+)\;,\quad
(D^c_1,D^c_2,D^c_3)=
(+,-,+)\;.
\end{align}
\item Type II-b,
\begin{align}
\left((Q_1,U^c_1),(Q_2,U^c_2),(Q_3,U^c_3)\right)&=
(+,-,+)\;,\quad
(D^c_1,D^c_2,D^c_3)=
(-,-,+)\;.
\end{align}
\end{itemize}

Type I-a gives the mass matrices as
\begin{align}
M_u(\text{I-a})=\begin{pmatrix}
B_6\lambda^6&B_4\lambda^5&RB_2\lambda^3\\
B_4\lambda^5&B_4\lambda^4&R\lambda^2\\
RB_2\lambda^3&R\lambda^2&1
\end{pmatrix}\alpha_U\langle H_u\rangle\;,\quad
M_d(\text{I-a})=\begin{pmatrix}
B_4\lambda^5&B_4\lambda^4&R\lambda^2\\
B_4\lambda^4&B_2\lambda^3&R\lambda\\
RB_2\lambda^4&RB_2\lambda^3&\lambda
\end{pmatrix}\alpha_D\langle H_d\rangle\;,
\end{align}
where $B_{2n}=\sum_{k=0}^{n}R^{2k}$.
Diagonalizing above matrices, we obtain mass ratios $m_c^2/m_t^2$, $m_s^2/m_b^2$, the CKM mixing angles (absolute values of CKM matrix elements), and the Kobayashi-Maskawa phase $\phi_3\equiv \arg(V_{ub}^*V_{ud}/V_{cb}^*V_{cd})$ at the leading order as
\begin{align}
&\frac{m_c^2}{m_t^2}=|1+R^4|^2\lambda^8\;,\quad
\frac{m_s^2}{m_b^2}=\frac{|1-R^4|^2}{(1+|R|^2)^2}\lambda^4\;,\nonumber\\
&\left|V_{us}\right|=\frac{2}{\sqrt{|R|^8+|R|^{-8}-2(2\cos^2 4\alpha 
-1)}}\lambda\;,\nonumber\\
&\left|V_{ub}\right|=
\frac{|R|||R|-|R|^{-1}|}{(|R|+|R|^{-1})\sqrt{|R|^4+|R|^{-4}+2
\cos 4\alpha}}\lambda^3\;,
\nonumber\\
&\left|V_{cb}\right|=
\frac{|R|\sqrt{|R|^2+|R|^{-2}+2\cos 4\alpha}}{|R|+|R|^{-1}}\lambda^2\;,
\label{CKM-TypeI}
\\
&\phi_1=\arg\left\{|R|^4+\frac1{|R|^2}+(|R|^2+1)\cos 4\alpha 
    -i(|R|^2-1)\sin 4\alpha \right\}\;,\nonumber\\
&\phi_2=\arg\left\{(1-|R|^2)\left(\frac{1}{|R|^4}-|R|^4
+2 i \sin 4\alpha\right)\right\}\;,
\nonumber\\
&\phi_3=
\arg\left\{(1-|R|^2)\left(|R|^4-\frac1{|R|^2}+(|R|^2-1)\cos 4\alpha 
    +i(|R|^2+1)\sin 4\alpha \right)\right\}.\nonumber
\end{align}
Let us discuss appropriate values of $|R|$, $\cos 4\alpha$ and $\sin 4\alpha$. We first expect that $\lambda\sim \sin\theta_c$. Then, $|R|> 1$ is needed to obtain the reasonable value of $|V_{ub}|$. However, $|R|$ cannot be much greater than unity, because $m_c^2/m_t^2$ exceeds the experimentally acceptable value. In addition, $\cos 4\alpha<0$ is necessary for $|R|>1$ to explain the data of $|V_{cb}|$. Finally, $\sin 4\alpha<0$ is required for $\phi_1$ to be in the first quadrant. In this case, however, it turns out that $\phi_3$ cannot be in the first quadrant simultaneously. 

For numerical evaluation, we take $|R|=\sqrt{3/2}$, $\cos 4\alpha=-3/4$ and $\sin 4\alpha=-\sqrt{7}/4$. This parameter set determines $\lambda=0.25$ under the experimental value of $|V_{us}|(=0.22)$. The matrices $M_{u,d}(\text{I-a})$ are diagonalized, and we obtain 
\begin{align}
|V_{ub}|=0.0028\;,\quad |V_{cb}|=0.032\;,
\end{align}
which are in excellent agreement with the CKM mixing angles at the GUT scale, $|V_{cb}|=0.029$--$0.039$ and $|V_{ub}|=0.0024$--$0.0038$ which are evaluated from renormalization group method with the experimental data at low energies\cite{das-parida}. However, quark mass ratios are predicted as
\begin{align}
\frac{m_c}{m_t}=0.0061\;,\quad
\frac{m_s}{m_b}=0.073\;,
\end{align}
which are about twice as large as the expected values at the GUT scale; 
$m_c/m_t\sim 0.0032\pm 0.0007$ and $m_s/m_b=0.036\pm 0.005$. Finally, we find 
\begin{equation}
\phi_1=12^\circ\;,\;\; \phi_2=31^\circ\;,\;\; \phi_3=137^\circ\; .
\end{equation}
Although the predictions cannot explain all the data simultaneously, it would be amazing to observe that this model can reproduce most of them in a considerable extent. We also find that for Type I-b, the mixing angles, mass ratios between 2nd- and 3rd-generations, and the CKM phase are completely the same at leading order as in Eq.~\eqref{CKM-TypeI}. 

For Type II-a, the mass matrices are
\begin{align}
M_u(\text{II-a})=\begin{pmatrix}
B_6\lambda^6&RB_4\lambda^5&B_2\lambda^3\\
RB_4\lambda^5&B_4\lambda^4&R\lambda^2\\
B_2\lambda^3&R\lambda^2&1
\end{pmatrix}\alpha_U\langle H_u\rangle\;,\quad
M_d(\text{II-a})=\begin{pmatrix}
B_4\lambda^5&RB_2\lambda^4&B_2\lambda^2\\
RB_2\lambda^4&B_2\lambda^3&R\lambda\\
B_4\lambda^4&RB_2\lambda^3&\lambda
\end{pmatrix}\alpha_D\langle H_d\rangle\;. 
\end{align}
The resulting mass ratios, the CKM parameters and phases $\phi_1$, $\phi_2$ and $\phi_3$ are the same as those in Type I except for $|V_{us}|$ and $|V_{ub}|$, which are 
\begin{align}
&\left|V_{us}\right|=
\frac{2|R|}{\sqrt{|R|^8+|R|^{-8}-2(2\cos^24\alpha-1)}}\lambda\;,\nonumber\\
&\left|V_{ub}\right|=\frac{|R|^2||R|-|R|^{-1}|}
{(|R|+|R|^{-1})\sqrt{|R|^4+|R|^{-4}+2\cos 4\alpha}}\lambda^3\;.
\end{align}
These expressions are different from those in Type I by the multiplication factor $|R|$. 

In this case, we take $|R|=\sqrt{3/2}$ and $\cos 4\alpha=-1/2$ in order to compensate the effect of the extra factor $|R|$ in $|V_{us}|$ in comparison with the that in Type I. In addition, we take $\sin 4\alpha=-\sqrt{3}/2$ and $\lambda=0.23$. We obtain the following results;
\begin{align}
&|V_{us}|=0.22\;,\quad |V_{ub}|=0.0038\;,\quad
|V_{cb}|=0.035\;,\nonumber\\
&\frac{m_c}{m_t}=0.0060\;,\quad \frac{m_s}{m_b}=0.059\;,\nonumber\\
&\phi_1=18^\circ  ,\; \phi_2=38^\circ  ,\; \phi_3=123^\circ.
\end{align}
Although the size of $m_s/m_b$ becomes smaller than that of the Type I, it is still too large to be phenomenologically acceptable. Moreover, $m_c/m_t$ remains two times greater than the expected value, and $\phi_3$ is in the 2nd quadrant. Type II-b gives almost the same results for the CKM parameters and the mass ratios between 2nd- and 3rd- generations.

\section{Conclusion}
We have studied possibility of incorporating CPV by using the FN mechanism in the context of democratic flavor FN couplings. We have considered models with two FN fields, in which the relative phase of their VEVs plays as the origin of CP violating phase at low energies. In the scenario with the $Z_2$ symmetry, the relationship among ratios between quark masses, the absolute values of CKM matrix elements and the CP violating phase has been examined in several simplest models. 
We have found that the predictions have been in good agreement with most of the data. However, the CP violating phase $\phi_3$ has been predicted to be around $130^\circ$, so that the models we have examined are not acceptable. It may be oversimplification to assume the flavor blind couplings. We expect that the small modification for the democratic assumption would cure this phenomenological problem. 

We have demonstrated the way how introduce the CPV to the FN model and have shown that the scenario with two FN fields would be promising. An application of our scenario to the lepton sector including neutrinos is under way and will appear in our future publications. If this will be successfully achieved, we would obtain a model which gives the unified description of CP phases for quarks and leptons. 

\vspace{1cm}
{\bf Acknowledgements.}  
S.~K. was supported, in part, by the Grant-in-Aid of                       
the Ministry of Education, Culture, Sports, Science and                         
Technology, Government of Japan, Grant Nos.17043008                             
and 18034004.
S.~P. and T.~S. were supported in part by the Italian MIUR and
INFN under the programs ``Fisica Astroparticellare''.


\end{document}